\documentstyle[aps,epsf,prl,multicol]{revtex}

\def\Journal#1#2#3#4{{#1} {\bf #2}, #3 (#4)}

\def\CQG{\em Class. Quantum Grav.}
\def\PRD{\em Phys. Rev. D }
\def\GRG{\em Gen. Rel. Grav.}

\newtheorem{theo}{Theorem}

\begin{document}
\title{Comment on ``Conformally flat stationary axisymmetric metrics''}
\author{Alan Barnes \\
School of Engineering \& Applied Sciences, Aston University,
Birmingham, B4 7ET, United Kingdom.}
\author{Jos\'e M. M. Senovilla \\
Departamento de F\'{\i}sica Te\'orica,  Universidad del Pa\'{\i}s Vasco,
Apartado 644, 48080 Bilbao, Spain.}

\maketitle
\begin{abstract}
In \cite{GC}, the authors claim to have found a previously overlooked 
family of stationary and axisymmetric conformally flat spacetimes, 
contradicting an old theorem of Collinson \cite{C}. In both \cite{GC}
and \cite{C} it is tacitly assumed that the isometry group is
orthogonally transitive. Under the same assumption, we point out here 
that Collinson's result still holds if one demands the existence of 
an axis of symmetry on which the axial Killing vector vanishes. 
On the other hand if the assumption of orthogonal
transitivity is dropped, a wider class of metrics is allowed and it is
possible to find explicit counterexamples to Collinson's result.
\end{abstract}

\begin{multicols}{2}
    In \cite{GC} the general form of conformally flat spacetimes 
admitting an Abelian 2-parameter orthogonally transitive group $G_{2}$
of isometries acting on timelike orbits is found. In that paper the
assumption of orthogonal transitivity is not mentioned explicitly, but
from the form of the line element given in equation (1) which Garc\'{\i}a
and Campuzano take as their starting point, it is clear that orthogonal
transitivity has been assumed.  The authors then prove that {\em not}
all those spacetimes can be diagonalized or, in other words, that  
the there are line-elements with no hypersurface-orthogonal timelike 
Killing vector field. Thus, these cases are properly stationary
(non-static) solutions. This result seems to contradict  
an important theorem due to Collinson \cite{C}, see also \cite{K,S}. 

However, Collinson's result still holds in an appropriate 
sense: if we require that the spacetime be truly axially symmetric
--- meaning that it contains a non-empty axis of symmetry ---, and 
not {\em merely} cyclically symmetric \cite{B}, then none of 
the new stationary metrics survive. Of course, every spacetime with a 
$G_{2}$ acting on timelike orbits can be made cyclically symmetric by simply 
closing the orbits of a spacelike Killing vector, that is, 
identifying two values of the appropriate coordinate. 
Nevertheless, this does not mean that the spacetime contains an axis 
of symmetry (the 2-dimensional cylinder is cyclically symmetric, but 
there is no axis: the axis is ``outside'' the manifold).
For a metric to have an axis it is necessary that the cyclic Killing vector 
vanishes on it, see e.g. \cite{MS} and references therein. As a matter 
of fact, this is independent of whether or not the axis is regular
\cite{MS}. If the axis is regular, then the elementary flatness 
condition must also be satisfied \cite{MS}. 

The cyclic Killing vector of the new stationary solutions found 
in \cite{GC}, formula (51), is given by $\partial_{\phi}$. A necessary 
condition so that $\partial_{\phi}$ vanishes on a would-be axis is 
that its scalar products with every other vector vanish there. This 
would require that --- using the notation of \cite{GC} --- either 
$e^{G(x, y)}=0$ somewhere, which is clearly not possible as the 
whole metric would be zero, or that $x=0$ {\em and} $1+x^2=0$, which is
also impossible. Thus, none of the stationary solutions found in 
\cite{GC} is axially symmetric, neither with a regular nor with a 
singular axis.

The conclusion is that Collinson theorem remains valid in the
orthogonally transitive case if one requires 
that the spacetime contains a non-empty axis. Combining the results 
in \cite{GC} and \cite{C}, we have
\begin{theo}
Every conformally flat stationary cyclically symmetric spacetime in
which the isometry group is orthogonally transitive, is 
given by the solutions in \cite{GC}. If in addition {\em axial} symmetry is 
required, then the spacetime is necessarily static. 
\end{theo}    

If the assumption of orthogonal transitivity is dropped then, as
pointed some years ago \cite{BR}, Collinson's result is not valid as
the following counterexample shows.
The Stephani metric \cite{St} with zero fluid expansion is:
$$ds^2 = (1+\frac{1}{4}Kr^2)^{-2}(dx^2 + dy^2 +dz^2 -V^2dt^2)$$
where $K=0,\pm 1$ and $V$ is given by
$$V= a(t)+b(t)r^2+{\bf r \cdot c}(t)$$
with $a$, $b$, $c_1$, $c_2$ and $c_3$ arbitrary functions of $t$. This
metric is a conformally flat perfect fluid solution (with fluid velocity
proportional to $\partial_t$).
The fluid energy density and pressure are given by 
$$\mu = -3K\qquad p=-3K+(aK+4b)(1+\frac{1}{4}Kr^2)/V$$
so that the density is constant, but the pressure normally depends on
all four coordinates.  In general the Stephani metric 
admits no isometries,
however, as shown in \cite{BR}, if we take $K = +1$,
$2a = 1-2A\sin(B t)$, $8b =1+2A\sin(B t)$, $c_1 = c_2 = 0$ and $c_3
=A \cos(B t)$ where $A$ and $B$ are non-zero constants, the metric admits a
(complete) two-dimensional isometry group with Killing vectors
$$\vec X =\partial_t 
+ B \left[\frac{1}{2}Kxz\partial_x
+\frac{1}{2}Kyz\partial_y+(1-\frac{1}{4}Kr^2+
\frac{1}{2}Kz^2)\partial_z\right]$$
$$\vec Y = x\partial_y - y\partial_x$$
This has an axis $x=y=0$ where the spacelike Killing vector $\vec Y$ vanishes and
the `tilted' Killing vector $\vec X$ is timelike at least in a region
around the origin $x=y=z=0$ if $B$ is sufficiently
small.  The vector $\vec X$ is not hypersurface orthogonal as may be
checked by direct calculation or by noting that if it {\it were}
hypersurface orthogonal it would be a Ricci eigenvector \cite{EK} which it
clearly is not (provided that $\mu+p \ne 0$). Furthermore note that the
fluid velocity vector does not lie in the two-plane spanned by $\vec X$ and
$\vec Y$ so that the fluid flow is convective; this is necessarily the case since, otherwise, the isometry group
would  be orthogonally transitive e.g. \cite{K}. Note also that there are similar
examples when $K=-1,0$; see \cite{BR} for full details.

The properly axially symmetric spacetimes include the 
physically relevant cases in which a fluid is to be matched to an exterior
spacetime across a spatially compact boundary, such as in the case of rotating 
stars. Observe also that the following classic statement \cite{C,K,S} 
is only true if the proviso of orthogonal transitivity is added: {\em The only conformally flat stationary and 
{\em axially} symmetric non-convective perfect fluid spacetime is the spherically symmetric
interior Schwarzschild solution with constant density}.

\end{multicols}

\end{document}